# Formation of MgB$_2$ at low temperatures by reaction of Mg with B$_6$Si

L.D. Cooley, Kyongha Kang, R. Klie, Q. Li, A. Moodenbaugh, and R. Sabatini

Materials Science Department, Brookhaven National Laboratory, Upton NY 11973

**Abstract**
Formation of MgB$_2$ by reactions of Mg with B$_6$Si and Mg with B were compared, the former also producing Mg$_2$Si as a major product. Compared to the binary system, the ternary reactions for identical time and temperature were more complete at 750 °C and below, as indicated by higher diamagnetic shielding and larger x-ray diffraction peak intensities relative to those of Mg. MgB$_2$ could be produced at temperatures as low as 450 °C by the ternary reaction. Analyses by electron microscopy, x-ray diffraction, and of the upper critical field show that Si does not enter the MgB$_2$ phase.

## 1. Introduction

Superconducting magnesium diboride may be well suited for electric power and magnet applications at temperatures below 30 K, due to its critical temperature $T_c$ of ~39 K and the abundance and low cost of Mg and B. Indeed, long-length wires and tapes of MgB$_2$ have already been fabricated by several routes [1], and very recently wound coils have been made with fields exceeding 1 tesla [2,3]. Additions of SiC nanoparticles to powder-in-tube wires have been shown to be especially potent at both increasing flux pinning and the upper critical field $H_{c2}$ [4]. However, the role of these additions is not clear. By remaining intact, the nanoparticles should add flux-pinning sites and limit grain growth, both factors being favorable for improved critical current density. By dissolving, on the other hand, Si and C might alloy with MgB$_2$ and produce increased electron scattering and higher $H_{c2}$ [5,6]. In the latter case, it has been speculated that the strong reduction of critical temperature due to carbon doping [7] might be offset by simultaneous silicon doping [6]. Nanostructural studies are so far inconclusive, due to the large variety of defects produced when Mg, B, and SiC react.

While the role of carbon is rapidly being clarified as a potent way to add electron scattering [8,9,10], much less is known about the effects of adding Si. Earlier attempts to react Mg, B, and Si powders suggested that Si clusters add flux pinning sites, but otherwise produce little if any change to the structure or superconducting properties of MgB$_2$ [11]. The addition of ZrSi$_2$ and WSi$_2$ also improved flux pinning but did not seem to alter the MgB$_2$ itself [12]. Additions of SiC and SiO$_2$ suggested that Si itself was not as important as the presence of C or O [13].

In this report, we help clarify the role of Si by investigating the ternary reaction 5Mg + B$_6$Si → 3MgB$_2$ + Mg$_2$Si using solid-state synthesis at 450 °C to 950 °C. Much more complete reactions were obtained by this route, especially at low temperatures, than for identical binary synthesis reactions Mg + 2B → MgB$_2$. This could be important for making *in-situ* wires where low heat-treatment temperatures and added pinning centers are desired. Superconducting properties of samples produced by either reaction were similar, suggesting that the ternary reaction does not incorporate Si into the MgB$_2$ phase. Electron microscopy and x-ray diffraction also found no evidence for Si incorporation into MgB$_2$.

## 2. Experiment

The samples used in this study were made by a single-step reaction, without intermediate grinding or refiring. Fine powders of Mg (99.8%, -325 mesh, Alfa Aesar), crystalline B (98%, -325 mesh, Alfa Aesar), and B$_6$Si (98%, -125 mesh, Alfa Aesar) were weighed and mixed in an argon glove box according to the reaction stoichiometry given above. Samples with 5 to 10 gram mass were then pressed uniaxially in a 9 mm die with approximately 20 kN force. Ductile flow of the Mg powder produced shiny, compact pellets as the result of this pressing. These pellets were wrapped in Ta or Mo foil and then placed in re-usable reaction crucibles made from type 316 stainless steel fittings (VCR type, Swagelok), which were sealed under 1 bar argon. A silver-plated stainless steel or nickel compression ring was used for the sealing surface between a compression nut and cap. Evidence of reaction between Mg vapor and the compression ring was observed at high temperatures, and trace amounts of magnesium chromate were also observed on the crucible after long reactions. Table I summarizes the reaction parameters and other characteristics of the experimental samples. The stainless steel containers were wrapped in stainless steel foil, placed into an open quartz tube, and inserted into a long tube furnace under flowing Ar5%H$_2$ gas. Samples were heated at 5 °C per minute and furnace cooled (~2 hours from 900 to 200 °C).

Table I. Reaction parameters and sample properties. "B" and "T" represent binary and ternary reactions, respectively. Shielding is expressed relative to perfect diamagnetism, and values above 100% indicate that void space is surrounded by loops of current.

| Sample | Temperature (°C) | Time (h) | $T_c$ (K) | $\Delta T_c$ (K) | Shielding | $d_{(100)}$ (Å) | $d_{(002)}$ (Å) | Unit cell volume (Å$^3$) |
|---|---|---|---|---|---|---|---|---|
| B1 | 550 | 100 | 38.2 | 22 | 8% | | | |
| B2 | 650 | 10 | 38.4 | 3.6 | 81% | 2.672 | 1.764 | 29.08 |
| B3 | 750 | 10 | 38.4 | 1.6 | 107% | 2.672 | 1.762 | 29.03 |
| B4 | 850 | 1 | 38.4 | 1.1 | 125% | 2.670 | 1.761 | 28.98 |
| B5 | 950 | 1 | 38.7 | 0.7 | 137% | 2.665 | 1.758 | 28.83 |
| T0 | 450 | 100 | 37.0 | 23 | <1% | | | |
| T1 | 550 | 100 | 36.8 | 4.8 | 86% | 2.673 | 1.756 | 28.96 |
| T2 | 650 | 10 | 36.7 | 2.0 | 113% | 2.660 | 1.758 | 28.73 |
| T3 | 750 | 10 | 38.0 | 1.3 | 107% | 2.665 | 1.758 | 28.82 |
| T4 | 850 | 1 | 38.0 | 1.3 | 94% | 2.669 | 1.761 | 28.97 |
| T5 | 950 | 1 | 37.5 | 2.7 | 96% | 2.663 | 1.758 | 28.80 |

Samples for experimental characterization were cut from the interior of the pellets. For magnetization measurements, long bars with ~50 mg mass were used, with the field applied parallel to the long axis. Measurements of the magnetic moment in a 1 mT field after cooling in zero field were made using a SQUID magnetometer. Pieces of these specimens or directly adjacent to these specimens were subsequently used for x-ray diffractometry and electron microscopy. X-ray powder diffraction data was acquired using copper $K_\alpha$ radiation using silicon as a reference. For transport measurements, thin bars were made by polishing to approximately 0.5 × 0.5 mm$^2$ cross section and 5 mm length. Silver paint was used to attach 50 μm diameter gold wires to these bars in the standard four-point configuration. Measurements were made as a function of field and temperature using a commercial measurement system (Quantum Design PPMS).

To explicitly search for Si inside the MgB$_2$ grains, Z-contrast imaging in scanning transmission electron microscopy mode (STEM) and correlated electron energy loss spectroscopy (EELS) [14] were used. These results were obtained using a JEOL-3000F STEM/TEM, equipped with a Schottky field-emission source operated at 300keV, an ultra high resolution objective lens pole piece (C$_s$=0.52 nm), and a post column Gatan imaging filter (GIF) for electron energy-loss spectroscopy. The microscope and GIF spectrometer were setup for a convergence angle (α) of 11 mrad and a spectrometer collection angle (θ$_c$) of 24 mrad, resulting in a probe size of 1.4Å. Z-contrast images are formed by collecting the high-angle scattering on an annular detector while the probe is scanned across the specimen, and the image intensity is approximately proportional to the average atomic number $Z^2$. The combination of Z-contrast imaging and EELS can be used to measure elemental concentration down to 0.2% [15].

## 3. Results and discussion

The magnetization data are shown in Fig. 1 and also summarized in Table I. The ternary reactions produced $T_c$ values of 36.7 to 38.0 K (Fig. 1a), somewhat lower than those for the binary reactions, which fell between 38.2 and 38.8 K (Fig. 1b). We speculate that the lower $T_c$ is the result of impurity atoms carried by the boron silicide, since it is known that the MgB$_2$ critical temperature is sensitive to boron purity [16]. $T_c$ became generally higher and sharper, as determined by the temperature difference $\Delta T_c$ between 10% and 90% of the magnetization at 5 K, for higher reaction temperature. For samples with full shielding (B2-B5, T1-T5), the shielding fraction presented in Table I is the percentage of perfect diamagnetism. In all, the binary series data is quite similar to that found by Rogado *et al* [17].

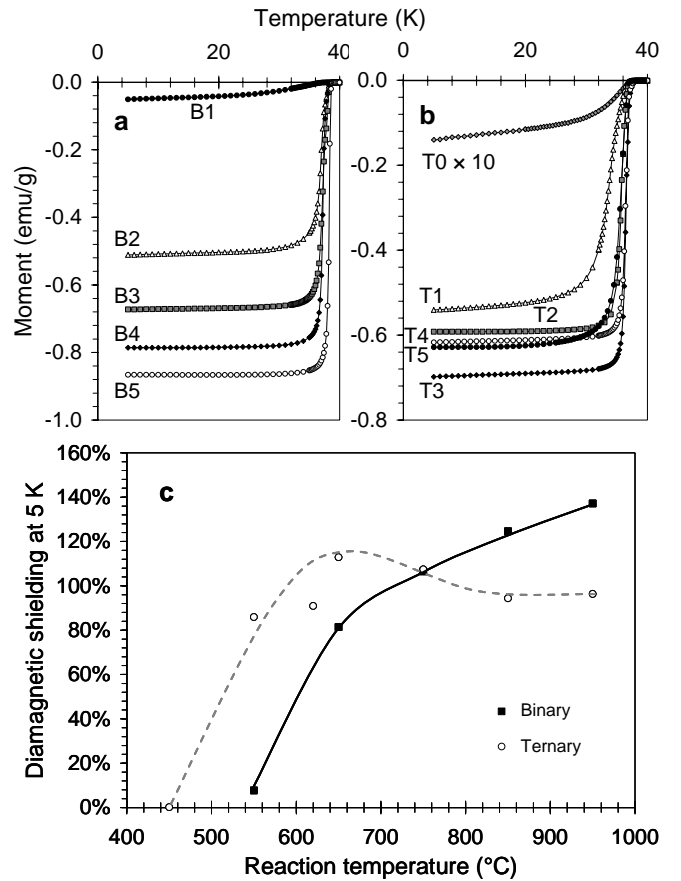

Fig. 1. Inductive superconducting transitions at 1 mT after cooling in zero field for samples made by the binary reaction (plot a) and the ternary reaction (plot b). The data for sample T0 in plot b has been multiplied by 10 to resolve it from the temperature axis. Plot c compares saturation magnetization as a fraction of diamagnetism for both sets of samples as a function of reaction temperature corresponding to the parameters in Table I.

The most striking difference between the ternary and binary reactions is the more rapid development of narrow magnetic transitions and large saturated moments as a function of increasing reaction temperature in the ternary samples. For instance, the ternary sample reacted at 550 °C (T1) produces a signal of -0.55 emu/g at 5 K, while the binary equivalent (B1) displays one-tenth of that signal. Most of these samples exhibit full diamagnetism, which we calculate to be 0.63 and 0.65 emu/g for the binary and ternary samples, respectively, based on the calculated density of the products (2.6 and 2.3 g/cm$^3$) and the volume of products relative to that of the reactants (76% and 84%). It should be noted that about 40% of $Mg_2Si$ by volume is produced in the ternary reaction. A more illustrative comparison is given in Fig. 1c. The ternary reactions produce full diamagnetic shielding at 550 °C and above, peaking at ~650 °C and falling somewhat at higher temperature. The binary reactions, by contrast, do not produce full shielding until 650 °C and above. Diamagnetic signals above 100% indicate that non-superconducting regions with low density are surrounded by magnetization currents, which could be voids or unreacted powders.

X-ray diffraction spectra are shown in Fig. 2. Only trace amounts of $MgB_2$ can be detected in sample B1 (Fig. 2a), and significant amounts of unreacted Mg are indicated for samples B2, B3, and B4. In comparison, more prominent $MgB_2$ peaks are indicated for sample T1 (Fig. 2b), and the Mg is exhausted in sample T4. Both B5 and T5 show slight amounts of MgO, while sample T5 also indicates substantial decomposition of $MgB_2$ by the many $MgB_4$ peaks (not labeled). On the whole, these spectra suggest more rapid conversion of reactants in the ternary sample. A more lucid demonstration of this point is given by the comparison of the intensities at 36.66° and 42.56°, corresponding to the locations of the (101) peaks for Mg and $MgB_2$, respectively, as shown in Fig. 2c.

Lattice parameters for $MgB_2$ were obtained by fitting the peaks of the x-ray diffraction patterns. The (100) and (002) lattice spacing and the unit cell volume so obtained are summarized in Table I. These analyses indicate that no systematic variation of the $MgB_2$ lattice is occurring as a result of the ternary reaction. If anything, a very slight expansion of the unit cell volume with decreasing reaction temperature occurred for the *binary* reaction, when compared to neutron diffraction analyses ($d_{(100)}$ = 2.668 Å; $d_{(002)}$ = 1.761 Å) [18]. This suggests that the ternary reaction does not incorporate Si into the $MgB_2$ phase.

Resistive transitions as a function of field $H$ and temperature $T$ are compared for samples B3 and T3 in Figs. 3a and 3b. These samples were almost completely reacted and also about 60% dense. As shown in Fig. 3, the ternary sample has substantially higher normal-state resistivity $\rho_N$ than the binary sample, 7.0 vs. 1.2 µΩ-cm. However, as shown in Fig. 3c there is little difference between the *H-T* curves corresponding to 90% of $\rho_N$, which can be identified as the parallel upper critical field $H_{c2//}(T)$ [19]. This further indicates that there is little if any Si alloying of $MgB_2$. The somewhat larger difference between the 10% $\rho_N$ curves may be due to obstruction of the current pathway by the large amount (~40% by volume) of $Mg_2Si$ that is formed. Scanning electron microscopy indicates that the $Mg_2Si$ grains have 0.1 to 10 μm typical diameters and are somewhat larger than the $MgB_2$ grains. Since these are too large to be effective as flux pinning centers, we surmise that the $Mg_2Si$ reduces the number of pathways connecting grains with **H** nearly parallel to the crystallographic *a-b* plane, leading to the lower 10% $\rho_N$ curve.

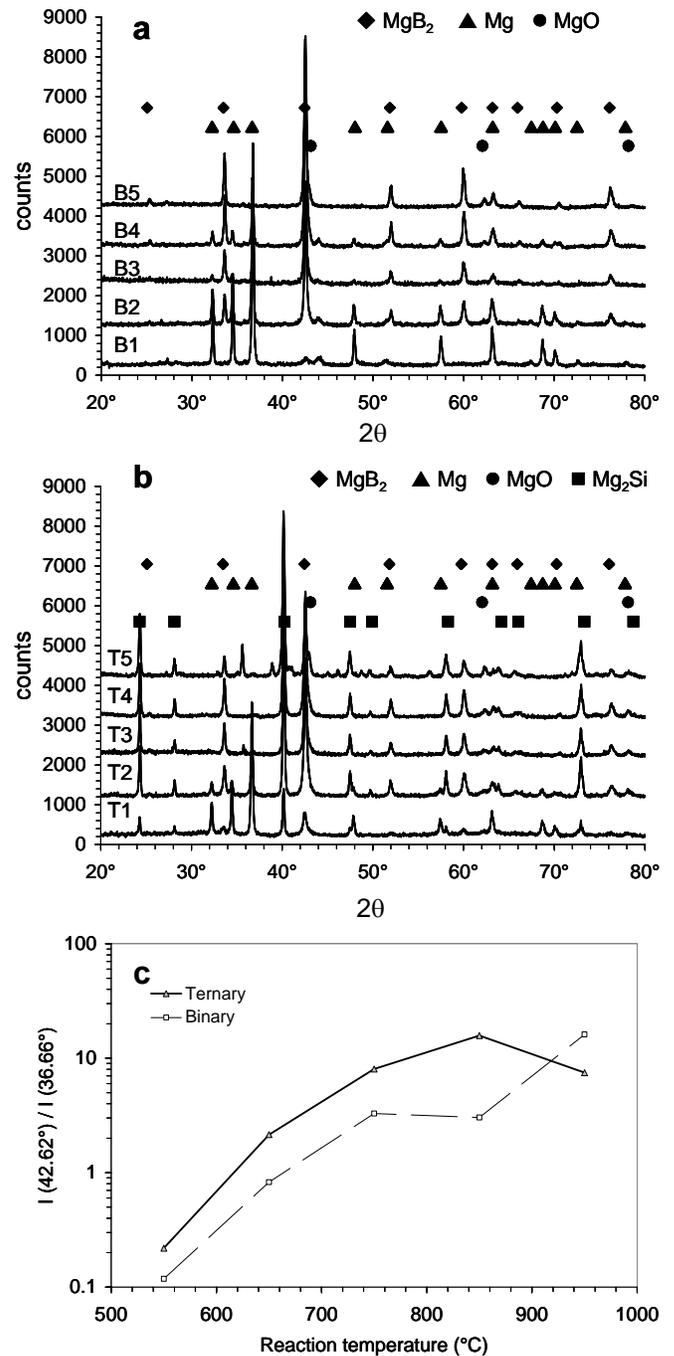

Fig. 2. X-ray diffraction patterns for samples made by the binary (plot a) and ternary (plot b) reactions. Peaks corresponding to the various phases present are indicated. Sample T5 also exhibits a number of peaks due to $MgB_4$, which are not labeled. Plot c shows the ratio of the intensity recorded at 42.62° to that at 36.66° as a function of reaction temperature (see Table I), which correspond to the (101) $MgB_2$ and Mg peaks, respectively.

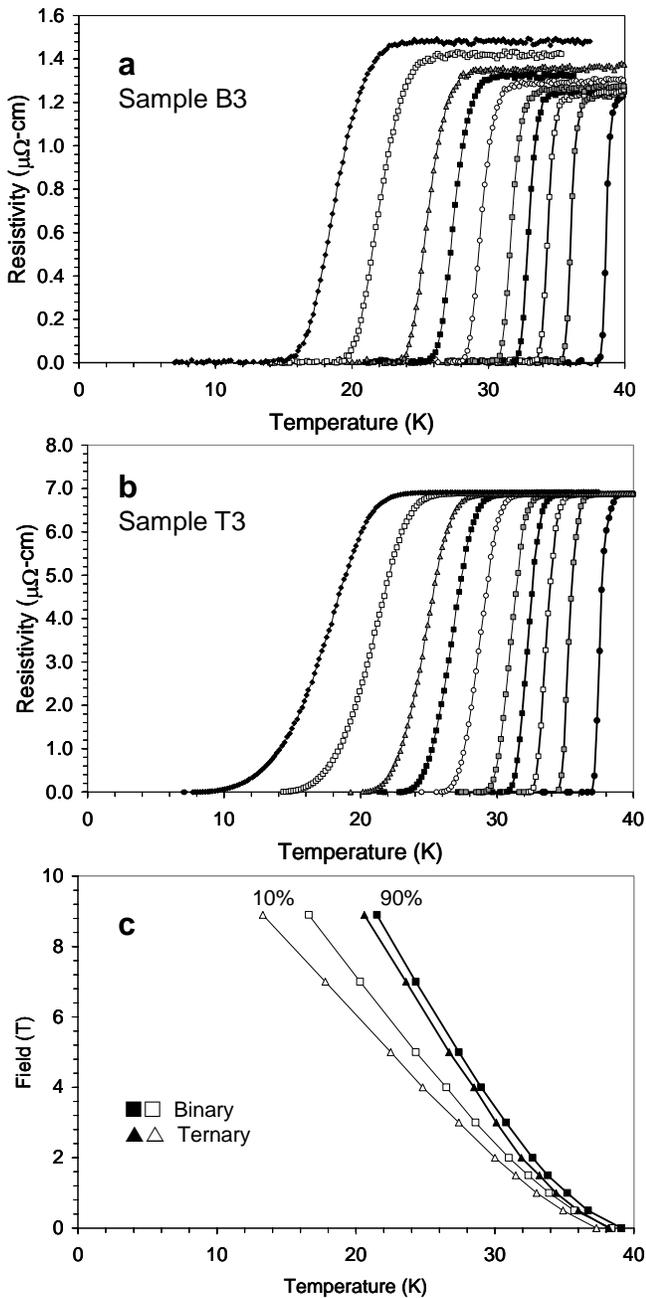

Fig. 3. Resistive transitions as a function of temperature in various magnetic fields for binary sample B3 (plot a) and ternary sample T3 (plot b). The fields are 0, 0.5, 1, 2, 3, 4, 5, 7, and 9 T for curves from right to left on both plots. Plot c shows the field-temperature data corresponding to 10% and 90% of the normal-state resistivity for the transitions in plots a and b.

Taken as a whole, the electromagnetic and x-ray diffraction investigations indicate that little if any silicon enters the $MgB_2$ phase. An explicit search for Si in sample T2 was conducted to verify this conclusion. Careful analysis over a wide range sample area revealed the three predominant phases: pure $MgB_2$, $Mg_2Si$, and unreacted $B_6Si$. This survey is consistent with the x-ray diffraction results, since orthorhombic $B_6Si$ produces few strong diffraction peaks and the sample preparation removes pure Mg. In addition, small amounts of oxide phases, such as $SiO_2$, $BO_x$, and $MgB_{2-x}O_x$, were incorporated inside some $MgB_2$ grains. These findings are consistent with past studies of impurity phases in $MgB_2$ [20]. However, no trace of Si was found inside $MgB_2$ grains. Most notably, the prominent EELS peak at about 100 eV, which corresponds to the Si-*L* edge, was clearly seen for $Mg_2Si$ grains but did not appear in any $MgB_2$ grain surveyed.

Based on these findings, it does not appear likely that Si contributes to doping of $MgB_2$ as suggested by previous reports where SiC nanoparticles where included in the reaction. In particular, the hypothesis [6] that Si might somehow offset the effects of carbon doping and prevent substantial $T_c$ decrease appears to not be valid. On the other hand, the copious production of $Mg_2Si$ might be beneficial for introducing flux-pinning centers. Indeed, SiC-doped samples likely receive two benefits—carbon doping to increase electron scattering and the upper critical field, and added nanoscale flux-pinning centers—when SiC nanoparticles are included as reactants.

The low temperatures at which $MgB_2$ can be formed might prove interesting for making wire composites. Presently, $Nb_3Sn$ composite superconductors undergo reactions anywhere from 48 hours at 675 °C for powder-in-tube designs to well over 200 hours at various temperatures and sometimes >100 hours at 650 °C for internal-tin designs [21]. The 100 hour heat treatment at 550 °C used for sample T1 is not excessive by these standards, and it is compatible with fiberglass-epoxy magnet insulation. If nanoscale mixtures of Mg and $B_6Si$ can be made, by high-energy ball milling for instance, these might also provide nanoscale flux-pinning centers after reaction. These possibilities deserve further examination, especially if an alloying element (perhaps Al) can be provided to increase the upper critical field.

## 4. Conclusions

The data we have shown demonstrate that $MgB_2$ forms more readily at low temperatures by the reaction of Mg with $B_6Si$ instead of with pure boron. The properties of the $MgB_2$ phase formed by the ternary reaction appear to be largely similar to those of samples formed by the binary reaction, suggesting that no alloying by Si takes place. Electron microscopy and x-ray diffraction also found no indication for Si alloying. Since a large amount, 40% by volume, of $Mg_2Si$ is formed as a by-product, the ternary reaction might be useful for forming flux pinning centers.

**Acknowledgments**

This work was supported by the U.S. Department of Energy, Division of Basic Energy Sciences under a BNL Laboratory-Directed R&D program. LDC would like to acknowledge discussions with V. Braccini, E. Hellstrom, D. Larbalestier, B. Senkowitz, and J. Waters at the University of Wisconsin.